**Altar: Structuring Sharable Experimental Data from Early Exploration to Publication**

William Gaultier*[1], Andrea Lodetti[2], Ian Coghill[3], David Colliaux[1], Maximilian Fleck[4], Aliénor Lahlou*[1,3]

[1] Paris Research, Sony Computer Science Laboratories, Paris, France
[2] Plant Sciences (IBG-2), Forschungszentrum Jülich GmbH, D-52425 Jülich, Germany
[3] CPCV, Department of Chemistry, École Normale Supérieure, PSL University, Sorbonne University, CNRS, Paris, France
[4] Institute of Chemistry for Life and Health Sciences, Chimie ParisTech, Université PSL, CNRS, Paris, France

Corresponding authors:

Willliam Gaultier, willi.gault@gmail.com
Aliénor Lahlou, alienor.lahlou@ens.psl.eu

Well-structured, carefully curated experimental data, arising from robust data management strategies, can have lasting impact, enabling breakthroughs years after their initial collection. For example, the Protein Data Bank (PDB) was essential to the training of AlphaFold. A wide range of published data management strategies exists to guide the creation of such datasets, with a strong emphasis on data publication for the broader scientific community and alignment with the FAIR principles (1). This late-stage focus can require retroactive reconstruction of data pipelines, increasing the risk of missing information or introducing errors (2). While necessary, it only partially addresses the broader scope of Data Management Plans (DMPs), which also define how data should be collected, documented, stored, and preserved throughout the project lifecycle, including its active development phase (3).

During the early stages of research projects, managing raw experimental data is challenging since experimental workflows evolve and data are often generated at high volume and speed. Resulting datasets tend to be large, heterogeneous, sparsely documented, accessible only on local computers and hard drives and tightly coupled to local execution environments. This makes them difficult to organize, interpret, or reuse by other project partners or even by the original experimenter over time. The rigid structure imposed by many existing data management strategies is often impractical in this context, as a project's underlying data model naturally evolves over the project. For example, in a chemistry project, compound characterization starts limited and broadens as the project evolves.

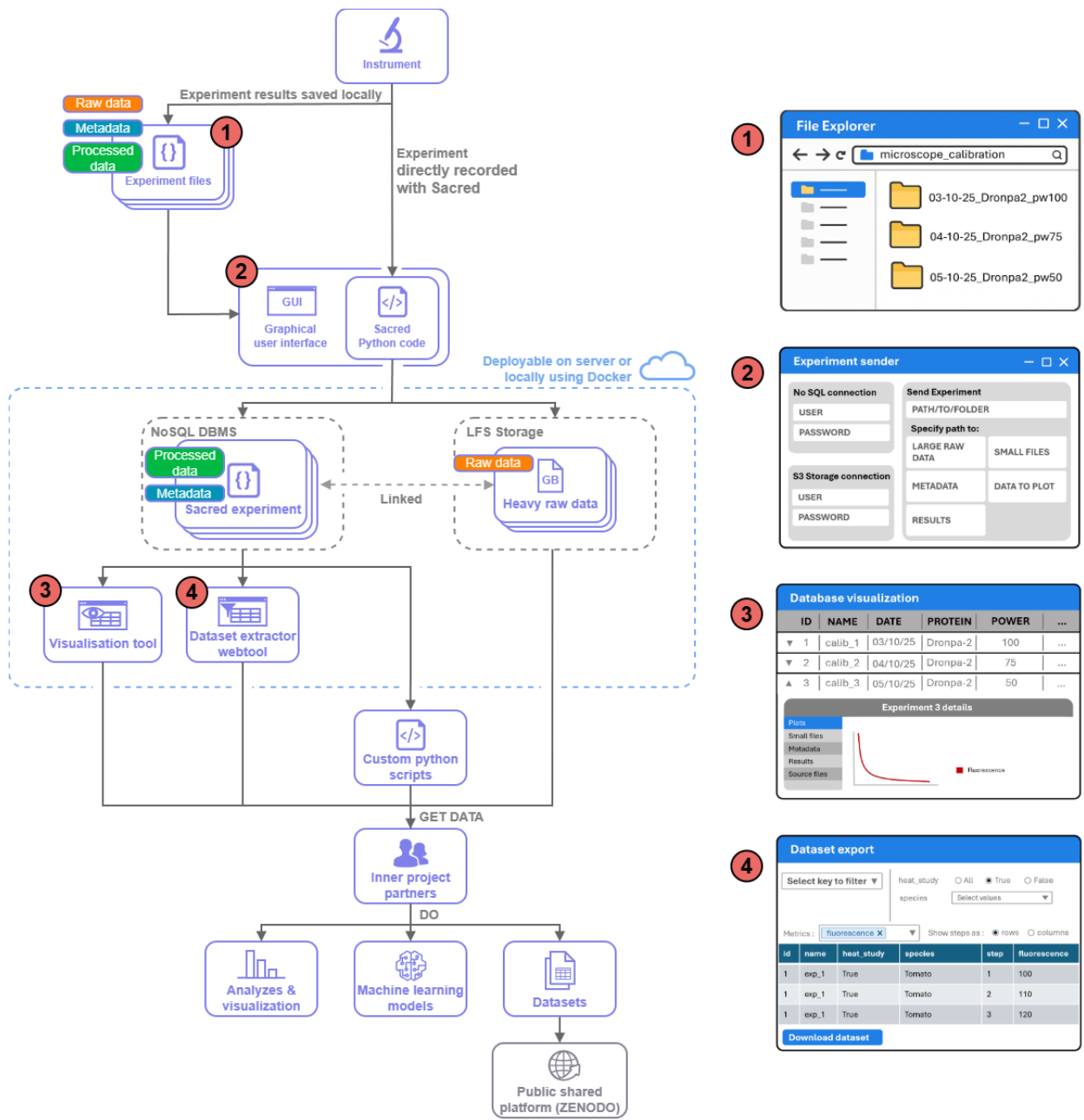

Figure 1: **Example of workflow articulated with the data management strategy in a collaborative project.** The experimental data can be obtained from the instrument and saved locally by the user (1), who then uses a graphical user interface (GUI - 2) to export them to the database. Alternatively, if the instrument can be controlled using Python, the data can automatically be sent to the database. The metadata, graphs and small files useful for visualization are saved in the NoSQL Database Management System (DBMS) for flexibility and quick query. The large raw files are saved in a Large File Storage (LFS) system and referenced in the database via a unique identifier (UID). The database content can be visualized (3) and downloaded either via a GUI (4) or Python scripts for post-processing and data curation and adaptation to communities' standards prior to dataset publication.

Efforts such as OMERO (4), a widely adopted server-based platform for managing microscopy data, illustrate how structured metadata handling, version control, and collaborative access can enhance scientific productivity, though OMERO remains a domain-specific tool optimized for imaging data. Electronic Laboratory Notebooks (ELNs) have also emerged as digital alternatives to paper lab notebooks, supporting experimental documentation and compliance (5). However, many classical ELNs lack formal data models and serve primarily as passive record-keeping tools, offering limited integration with broader data processing pipelines (6). This highlights a clear unmet need for flexible strategies suitable for the early stages of research, providing sufficient consistency to support collaboration while retaining the adaptability required during exploratory work. We adapted tools originally developed for machine learning to systematically track experimental configurations and data. This strategy is compatible with instrument-based workflows, allowing direct capture of raw data with minimal disruption or manual upload of experimental folders. The Altar suite provides an interface layer to implement the data management strategy, enabling flexible adoption across diverse research projects. The documentation of Altar allows users to adapt the strategy depending on their computer skillset (PhD students, post-docs, Principal Investigators, System Administrators). We provide in Supplementary Table S1 the set of tools selected in Altar, as well as alternatives.

The solution relies on the Sacred data model (Python) which was developed to track Machine Learning experiments (7). Experiments are saved following a strict data model which is highly flexible in practice: experimental parameters are saved as *configuration* metadata, plots are saved as *metrics*, files are saved as *artifacts*, and the original script, run logs, and host details are captured automatically. This unified approach ensures reproducibility and comprehensive documentation of each experiment. Experiments can be logged either by adding a few lines to existing Python scripts using the original Sacred framework (see Supplementary Information) or by transmitting locally generated data via the AltarSender application.

When an experiment is executed, a record is stored in a flexible NoSQL Database Management System (DBMS) that naturally handles heterogeneous, document-based data structures (8). This allows early and flexible data capture even when the final data model is not yet defined. While configuration, run details, data curves and small files are stored in the DBMS, Large raw data files are preferentially stored separately in a Large File Storage (LFS) system under a unique identifier (UID) (9). This avoids storage inefficiencies, write amplification, and interoperability issues while maintaining the link between raw data and the experimental record: the UID is referenced inside the database, and the experimental configuration is saved in the LFS to ensure bidirectional information retrieval. The database can be hosted on personal computers, on a server, or on the cloud. It can be migrated across platforms, ensuring that data remains organized, reproducible, and accessible to collaborators across different steps of the projects (details are provided in AltarDocker).

Once experiments are logged into the database via Sacred, users can access and interact with the stored data. Experiments can be explored and compared through web-based interfaces that allow users to inspect metadata, visualize plots, and download results or logs. We provide AltarViewer as an example. Data are accessible programmatically, and a non-coding option is provided via AltarExtractor for metric extraction.

The Altar framework has been already deployed to address common challenges in research projects: (i) use by individual researchers to organize locally data, simulations and models (ii) integration as data management apps for instrument users (10) (iii) extraction of dataset for publication on Zenodo (11) (iv) support for collaborative projects involving research partners across different domains.

The strategy implemented in Altar provides a flexible and practical framework for managing experimental data during the active phase of research. It captures both metadata and experimental data without imposing rigid data models beyond the Sacred model: configuration/metrics/files, stored in a NoSQL framework while large raw datasets are stored efficiently and linked to the database. The framework does not require a specialized infrastructure at the beginning of the project, yet can be easily scaled to the cloud and made publicly accessible when preparing data for collaboration and publication. The approach can enhance reproducibility, collaboration within project teams, and FAIR-aligned data sharing, supporting rigorous and shareable research across diverse scientific domains. Taken together, these features make the strategy well suited to be incorporated into a broader Data Management Plan (DMP). We provide a series of tools in the Altar suite to lower the entry barrier of this strategy.

**Acknowledgements**
The authors thank Shizue Matsubara and Marine Collery for useful discussions, and Jonathan Legrand, Ludovic Jullien, Michael Anslow for comments on the manuscript. The authors thank users from Sony CSL Paris, ENS-PSL, IBPC, FZJ, TUE and UPOL for testing the different Altar interfaces and the deployment documentation. The authors thank Pratik Bhoir and Samir Dhane for technical support and Sébastien Marino for illustrations.

**Fundings**
European Innovation Council Pathfinder Open DREAM (grant no. 101046451)

**Author Contributions**
WG, AL was involved in conceptualization. WG, DC, AL were involved in methodology. WG, AL were involved in software. WG, ALo, IC, DC, MF, AL were involved in resources and investigation. WG, DC, MF, AL were involved in writing – original draft. WG, ALo, IC, DC, MF, AL were involved in writing. WG, AL were involved in documenting and testing. AL was involved in supervision.

**Competing interests**
The authors declare no competing interests.

**Code availability**
The code is archived on Zenodo (10.5281/zenodo.18591058), published on Github (https://github.com/DreamRepo/Altar), documented on Github Pages (https://dreamrepo.github.io/Altar/) and a simple demo of the database visualisation is available here: https://altar-demo.cslparis.eu/sacred (password protected - contact the authors for access).

*Table S1: Tools available to implement the data management strategy. In bold, tools implemented in the Altar Suite.*

| Objective | Possible solutions | Altar tools |
|---|---|---|
| Data model | **Sacred** | AltarSender |
| Database management system | File Storage, **MongoDB**, TinyDB, SQL, S3 | AltarSender |
| Large File Storage (files >25MB) | **MinIO**, Ceph, s3gw | AltarSender |
| Database viewer | SacredBoard, **Omniboard**, **PyGwalker** | AltarViewer, AltarExtractor |
| Database extractor | **Incense**, PyMongo | AltarExtractor |

**Experiment Tracking with Sacred**

The Open-Source framework Sacred[1], implemented in Python, was initially developed to track machine learning framework optimisation experiments, and in particular training with different hyperparameters. Here we applied it for hardware-based experiments.

Sacred provides an efficient and unified data model for managing experimental metadata such as parameters, results, experimental code, versions and dependencies, experimental logs, and run identifiers in a consistent and structured way. This abstraction simplifies the handling, storage, and sharing of experimental data. Although other tools like MLflow [2] and Weights & Biases (wandb) [3] offer more specialized features such as dashboards, model versioning, and cloud collaboration, we opted for Sacred because of its simplicity, low overhead and open-source nature. An important aspect of Sacred, which is pertinent to scientific research, is that an experiment that has been run with Sacred is accessible but not easily editable to ensure immutability and avoid alteration of the original data and metadata[4]. To store these Sacred experiments and their associated metadata, we use MongoDB, a flexible NoSQL database well-suited for heterogeneous scientific data. The data can be uploaded from already saved folders, and AlterSender provides an example of interface. The data can also be uploaded directly at the experimental stage provided the instrument can be controlled through an interface for automation such as Serial communication, a Network protocol, a DLL, an API, a command-line interface or a Python library (see code snippet Figure S1).

**Centralized Storage with MongoDB**

The Sacred experiments are stored using MongoDB, an open-source, free-to-use NoSQL database that naturally handles heterogeneous, document-based data structures, making it well suited for storing the varied outputs of scientific experiments. NoSQL databases are flexible because they do not rely on a predefined structure: input fields can vary across experiments and missing values present no issue[5]. This approach enables early and flexible data capture, even when the final data structure is not yet defined. By encouraging researchers to log as much experimental context and metadata as possible from the earliest exploration stages, we reduce the risk of information loss and allow meaningful structure to emerge later through post-processing and annotation. Its flexibility allows experiment entries, each including configurations, results, logs, and metadata, to be stored as self-contained documents. MongoDB can be easily set up on a shared server, enabling all collaborators to access and query experimental data through a unified interface. This setup ensures consistency and facilitates collaboration. Importantly, the database can be hosted and moved between local storage, local server, shared server, or cloud-managed server locations, depending on the desired level of data circulation.

**Handling Large Raw Data**

One limitation of Sacred is its inability to handle large raw data files, such as video recordings from experiments. Sacred is primarily designed for logging configurations, metrics, and metadata, and is not well suited for storing or transferring heavy binary data, for which file storage systems are more adapted. Recent research confirms that storing large binary objects directly in databases often leads to storage inefficiencies, increased write amplification, and interoperability issues with external tools. While some attempts have been introduced to combine both[6], they require significant database engineering skills, which are rarely available in typical research projects. It further supports our choice to decouple raw data storage from experimental metadata, compatible with the FAIR principle that metadata should remain accessible even when the raw data are no longer available[7]. We set up a separate large file storage system using MinIO[8], an open-source, high-performance object storage solution compatible with the Amazon S3 API. MinIO is lightweight, easy, and fast to deploy, making it particularly well suited for research environments where rapid setup and minimal maintenance are important. Raw experimental data are stored externally in MinIO where each file has a unique identifier and is linked by a URL reference within the Sacred database[7]. This setup ensures a seamless connection between metadata and raw data, while preserving Sacred's lightweight architecture and avoiding the strain of managing large binary files directly within the system. It also limits unnecessary duplication of those large files when migrating Sacred databases. This decoupling of metadata (in MongoDB) from binary payloads (in MinIO) ensures scalable and efficient storage and retrieval.

```python
from sacred.observers import MongoObserver
from sacred import Experiment
ex = Experiment("get_movie")
ex.observers.append(MongoObserver(db_name='demo'))

import PulseGenerator
import Camera

@ex.config
def config():
    exp_duration  = 100 #s
    frame_acquisition = {frame_rate = 10, #Hz,
                         exposure = 100, #ms
                         gain = 10}
    pulse = {pulse_frequency = 1, #Hz
             pulse_delay = 3, #s
             pulse_amplitude = 10, #mA
             LED_pin = 1}
    plant_species = "Arabidopsis thaliana"
    port_camera = "COM3"
    port_control = "COM6"
    save_folder = "demo/"

@ex.automain
def run(_run, port_control, port_camera, pulse, frame_acquisition,
    save_folder):
    control = PulseGenerator.connect(port_control)
    # LED excitation
    control.create_pulses(pulse)
    # Camera settings
    camera = Camera.connect(port_camera)
    camera.update_settings(frame_acquisition)
    # run experiment
    camera.wait_for_trigger()
    control.start()
    video = camera.video
    control.disconnect()
    camera.disconnect()
    # record fluorescence
    for i, frame in enumerate(video.frames):
        _run.log_scalar("Average fluorescence", frame.mean(), i/
            frame_acquisition['frame_rate'])
```

*Figure S1: Example of instrument automation with Python and record of an experiment with Sacred. Here a camera acquires a movie of the response (fluorescence) of a sample (plant Arabidopsis thaliana) exposed to a pulsed light signal from an LED. The camera and pulse generator are controlled via Python. For simplicity quick visualization in the database, the average image of the movie is recorded in the database (see section on handling large raw data for video recording).*

**Accessing and Exploring the Data**

Once experiments are logged into MongoDB via Sacred, users can access and interact with the stored data using different tools depending on their needs. For visual exploration and comparison of experiment runs, we use Omniboard[9], a web-based dashboard that connects to the MongoDB database and allows users to browse experiments, see metadata, observe data plots, and view or download results, logs or files. For more programmatic access, especially when integrating experimental data into analysis pipelines, we rely on Incense[10], a lightweight Python library designed to query Sacred experimental data directly from MongoDB. The process is illustrated in Figure S2, which shows how Incense can be used to collect experimental data from a Sacred database. We also used Incense to create a dedicated web tool (AltarExtractor) for exploring experiments with metadata filtering, and to access related raw data or experiment results. This dual approach ensures that both technical and non-technical users can efficiently access, explore, and reuse experimental results and their associated metadata.

```python
from incense import ExperimentLoader
from custom_tools import extract_metadata_from_config,
    process_experiment_run

# Step 1: Connect to the Sacred MongoDB database using Incense
loader = ExperimentLoader(mongo_uri="mongodb://host:port", db_name=
    "sacred")

# Step 2: Query the desired experiment runs
query = {"experiment.name": "EXPERIMENT_NAME"}
runs = loader.find(query)

# Step 4: Iterate over runs and extract metadata, video link, and
    metrics
for run in runs:
    config = run.config
    metrics = run.metrics
    video_link = config.get("video_link")
    metadata = extract_metadata_from_config(config)

    # Step 5: Use the data (e.g., display, aggregate, analyze)
    process_experiment_run(metadata, metrics, video_link)
```

*Figure S2: Python code snippet showing how Incense is used to get data from a Sacred database.*